# Feynnman Diagram Calculations

# From finite Integral Representations to knotted Infinities [*]

Dirk Kreimer[†‡]
*Department of Physics, University of Mainz*
*55099 Mainz, Germany*

ABSTRACT

Work is reported on finite integral representations for 2-loop massive 2-, 3- and 4-point functions, using orthogonal and parallel space variables. It is shown that this can be utilized to cover particles with arbitrary spin (tensor integrals), and that UV divergences can be absorbed in an algebraic manner. This includes a classification of UV divergences by means of the topology of the graph, interpreted in terms of knots.

## 1. Introduction

The following sections will cover three topics: Section 2 discusses the use of parallel and orthogonal space variables. It is shown how to simplify the $\gamma$-algebra, and how to construct what should be called the characteristic polynomial of a graph. Section 3 reports on work done in collaboration with the theory group at Mainz university. It focuses on algorithms written in Maple, heading towards a complete algebraization of the scheme outlined in Section 2. We also mention some algebraic properties of the UV sector of a renormalizable field theory. Section 4 focuses on recent calculations with David Broadhurst, classifying various Feynman graphs with the help of knot theory.

## 2. Parallel and Orthogonal Space Variables

It is one of the major challenges in computational high energy physics to obtain results for 2-, 3- and 4-point functions at the two-loop level involving various internal masses. Quite often it turns out that such results can be obtained using orthogonal and parallel space variables. In the following we recapitulate the ideas behind the introduction of these variables and, in particular, explore how they might be used to facilitate the handling of the $\gamma$-algebra.

Assume we are confronted with the calculation of a $n$-point function in perturbative quantum field theory. The exterior momenta $p_j$ span a $n-1$ dimensional vector space, due to momentum conservation. Necessarily, all loop momenta $l_i$ involved

---

split in a natural way into two components:

$$l_i^\mu \to l_{i,=}^\mu + l_{i,\perp}^\mu,$$
$$l_{i,\perp} \cdot p_j = 0 \quad \forall \ i,j.$$

For loop integrations this defines a set of parallel space variables $l_{i,=}$ and orthogonal space variables $l_{i,\perp}$ in a covariant manner; $l_{i,=}^\mu$ and $l_{i,\perp}^\mu$ are genuine four-vectors.

Dimensional regularization uses this splitting to allow for an operational definition in terms of analytic continuation of the radial orthogonal space variables.[1] Much of our previous work relies on utilizing this set of variables not only to prove various properties of DR, as done in Collins' book, but to simplify actual calculations.[1,2,3] One can obtain useful integral representations, including even the four-point two-loop box topologies, along the following lines:[2,3,4,5,6]

- Choose a measure $\int_{-\infty}^{\infty}$ for parallel space variables.

- Integrate simply the angular integrations in the orthogonal space.

- Integrate the modulus of the four-vectors $l_{i,\perp}$.

- Do as many as possible parallel space integrations via the residue theorem.

In particular, the method can be extended to obtain integral representations for a whole graph, and we will comment on these ideas in more detail now.[5] Take for example a two-loop calculation of a fermionic vertex correction. We have a two-dimensional parallel space, and the vertex carries a Lorentz index $\alpha$ say. Let us describe how we can obtain an integral representation for such a problem.

Without loss of generality we can choose two vectors $e_1^\mu, e_2^\mu$, $e_1 \perp e_2 = 0$ so that

$$\begin{aligned} p_1^\mu &= p_{11} e_1^\mu + p_{12} e_2^\mu, \\ p_2^\mu &= p_{21} e_1^\mu + p_{22} e_2^\mu, \\ l_1^\mu &= l_{11} e_1^\mu + l_{12} e_2^\mu + l_{1,\perp}^\mu, \\ l_2^\mu &= l_{21} e_1^\mu + l_{22} e_2^\mu + l_{2,\perp}^\mu, \end{aligned}$$

and we use a measure for loop integrations

$$\int_{-\infty}^{+\infty} dl_{11} dl_{12} dl_{21} dl_{22} \int_0^{+\infty} d|l_{1,\perp}|\, d|l_{2,\perp}| \int_{-1}^{+1} d(l_{1,\perp} \cdot l_{2,\perp}).$$

In the numerator we generally come across a lengthy string of $\gamma$-matrices. We could go the traditional way, trace it, project out the various form factors, and calculate all scalar integrals which were obtained. Such a method was featured by the Leiden group.[7] Here we will advocate a different approach, as introduced recently.[5] We like to utilize some elementary properties of the $\gamma$-algebra, combined with our parallel/orthogonal space variables. Corresponding to $e_1, e_2$ we find imbedded in the $\gamma$-algebra elements $\gamma_1, \gamma_2, \gamma_\perp^i$ so that:[5]

$$\{\gamma_1, \gamma_2\} = \{\gamma_1, \gamma_\perp^i\} = \{\gamma_2, \gamma_\perp^i\} = 0,$$
$$\{\gamma_\perp^i, \gamma_\perp^j\} = 2g_\perp^{ij},$$
$$g_{\perp i}^i = D - 2,$$
$$\gamma_1^2 = 1, \gamma_2^2 = -1, \gamma_{\perp,i}^2 = -1. \tag{1}$$

Here the matrices $\gamma_{\perp,i}$ forms a basis in orthogonal space, and we have assumed without loss of generality that the basis vector $e_1$ is timelike. We have

$$\slashed{l}_i = l_{i1}\gamma_1 + l_{i2}\gamma_2 + l_\perp \cdot \gamma_\perp. \tag{2}$$

A crucial observation is now that we can replace the whole $\gamma$-algebra in the orthogonal space by a totally symmetrized combination of metrical tensors $g_\perp$. This is because these tensors deliver the only possible formfactors in the orthogonal space. No exterior momenta can modify them. We further note that the external index $\alpha$ has to be taken into account by using $\gamma^\alpha = (\gamma_1, \gamma_2, \gamma_\perp^i)$.

Utilizing Eq.(1) and Eq.(2) we see that the $\gamma$-algebra reduces to a sum of terms

$$f_{n_5, n_1, n_2} \gamma_5^{n_5} \gamma_1^{n_1} \gamma_2^{n_2},$$
$$n_5, n_1, n_2 \in \{0, 1\}, \tag{3}$$

with scalar functions $f_{n_5, n_1, n_2}$.

Here we incorporated a possible $\gamma_5$, which can be treated either as anticommuting, as justified recently, or along the more cumbersome BM approach.[8,9,10] Note that even with a non-anticommuting $\gamma_5$ we have

$$\{\gamma_5, \gamma_1\} = \{\gamma_5, \gamma_2\} = 0,$$

as the nonvanishing part of the anticommutator lives in orthogonal space. These are the advantages: We can obtain an answer without applying the trace to our amplitude. If we are interested in a trace we may apply it at the end, where it becomes trivial when acting on the terms of Eq.(3). The whole method can be implemented by using non-commutative operators in REDUCE for example. In doing so we found that this way of handling the $\gamma$-algebra is superior to the standard ways incorporating lengthy 'High Energy Packages'. This article is not the place to present elaborate examples, but we intend to incorporate the experience gained at the one-loop level also in the two-loop level and will report on the results in the future.[11]

## 3. UV-Divergences and Algebraic Structures

To exploit the advantages of the method presented in section 2 one has to isolate the UV singular behaviour of the theory. This involves not only genuine two-loop contributions, but also various one-loop terms. Based on earlier results a one-loop package is now available which facilitates one- and two-loop calculations.[12,13,14,15] It is presented in a form which makes it useful for applications which even advocate a more traditional approach to two-loop calculations. Here we shortly comment on some improvements of the package compared to the 1.0 version published recently.[15]

- Terms linear in $(D-4)$, useful for two-loop calculations, are now evaluated as well.

- Degenerate cases for exterior momenta and masses can now be handled.

- The program is now able to treat IR divergent cases.

A similar package for the two-loop level is obtained for two-point functions but has yet to be tested. The package follows the ideas outlined in section 2, isolates UV-singularities as described elsewhere, incorporating the necessary counterterm graphs during this process, and therefore utilizing at a modest two-loop level some remarkable structures of UV divergent Feynman graphs.[5] In fact, it is shown that the UV-divergent structure of a Feynman graph can be obtained in an algebraic manner.[16,17] For ladder topologies a full algebraization can be achieved, for nested as well as overlapping divergences. In the following section we describe the ideas which were used to obtain a topological classification of UV-divergences.

## 4. Classifications of Feynman Graphs

The study of UV-divergences in Feynman graphs reveals some remarkable properties: topologically simple graphs (ladder graphs) are free of transcendental contributions in the divergent sector, and the topological nature of more complicated graphs reflects itself in the transcendental contributions to $Z$-factors.[16,17] This pattern has been confirmed by studying various results of Broadhurst.[18] These results have motivated David Broadhurst to investigate all Feynman graphs which correspond to the vertex correction at the seven loop level in $\phi^4$ theory, and which are free of subdivergences.[19] In collaboration with him, the author started a classification of the Feynman diagrams in terms of knots, elaborating on the ideas developed in previous work.[16,17] Details of the calculation will be presented elsewhere; some of the results can also be found in Broadhurst's contribution to these proceedings.[19,20]

We now describe the steps necessary to classify Feynman diagrams using knot theory.

- First we map the $\phi^4$ couplings to three-point couplings in all possible ways. The intermediate propagator is a static propagator corresponding to a Lagrange-multiplier field.[§]

- Then we map the Feynman graph to a link diagram following the rules of as given in previous work.[17].

---

[§]We thank Bob Delbourgo for supporting us with this idea to map four-point couplings to three-point couplings.

- We use a skein relation to reduce the link diagram to knots and try to relate the result to the transcendentals obtained by Broadhurst.

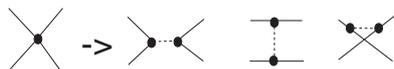

Fig. 1. The mapping of a four-point to three-point couplings in all possible ways. The dashed line indicates the static propagator.

We give an example borrowed from our previous work.[17]

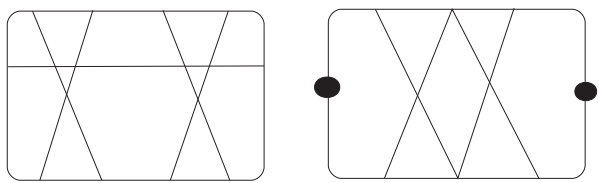

Fig. 2. We like to investigate this six-loop Feynman graph. We also give a $\phi^4$ graph which is topologically equivalent, and which was investigated by Broadhurst.[18] The two dots in this graph have to be identified. It can be obtained from the graph on the lhs by shrinking three propagators.

Let us map this graph to a link diagram:

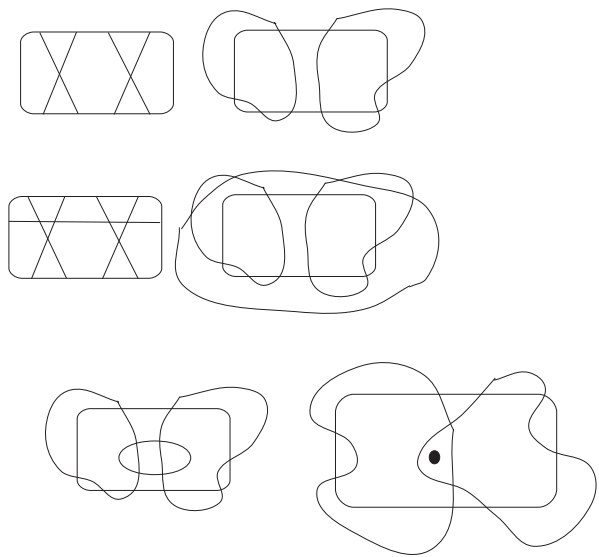

Fig. 3. The generation of the $(3,4)$ torus knot. In the first line we have removed one propagator to generate the $\zeta(3)\zeta(3)$ factor knot. Then we attach the last propagator in the most economic way, giving us the link diagram on the bottom rhs. We used Reidemeister II and III moves to get from the second to the third line. We end up with the braid word $\sigma_1^4\sigma_2\sigma_1^4\sigma_2$. (All components encircle the dot in the middle counterclockwise, so that we can read off the braid word.) After skeining the two kidneys we find a knot. It can be identified as the $8_{19}$ knot in the standard tables, which is the $(3,4)$ torus knot.[21]

Following these lines we hope that in future work an identification of the UV-divergent sector of Feynman graphs can be made possible from their topology.

**Acknowledgements**

It is a pleasure to thank the group in Mainz, especially Lars Brücher, Johannes Franzkowski, Ulrich Kilian, Jürgen Körner and Karl Schilcher for a long lasting collaboration on the subject of sections 2 and 3. The results sketched in section 4 are the consequence of an inspiring collaboration with David Broadhurst.

I would like to thank the organizers of this conference for their kind hospitality and support.

This work was supported under grant A69231484 from the Australian Research Council and under grant HUCAM CHRX-CT94-0579 from the EU.